\documentstyle[preprint,revtex]{aps}
\begin{document}
\draft
\preprint{Talk presented at STATPHY19, China, 1995}
\begin{title}
One Dimensional Lattice Models of Electrons\\with $r^{-2}$
Hopping and Exchange
\end{title}
\author{Ch. Gruber and D. F. Wang}
\begin{instit}
Institut de Physique Th\'eorique\\
\'Ecole Polytechnique F\'ed\'erale de Lausanne\\
PHB-Ecublens, CH-1015 Lausanne-Switzerland.
\end{instit}
\begin{abstract}
This talk reviews some recent progresses
in the study of low dimensional electron systems of strong
correlations.
\end{abstract}

\section{Introduction: Interest of 1D Systems}

The study of one dimensional systems is almost as old as
statistical mechanics. In the early days it might have been
considered to be just a ``warm up'' exercise necessary
before going to the study of the real 3-dimensional world.
However, in the last decades, the study of strongly correlated
electron systems has been very active to understand
general properties of condensed matter, such as metal-insulator
transition and high temperature superconductor, and 1D systems
are expected to throw some light on the mechanisms underlying
such phenomena. More recently, the discovery of high
temperature superconductivity, and the suggestion by Anderson
that the 2D Hubbard model could be appropriate to describe
these new materials, have stimulated considerable interest
in the study of electron systems in low dimensions. This is
due to the possibility that the normal state of these 2D
superconducting materials may share some properties of
the 1D interacting electron systems, the non-Fermi-liquid
behaviors. Finally, we wish to mention that today one can realize
1D conductors and 1D magnets experimentally.

\section{1D Lattice Models: A Short Review}

The systems usually considered consist of identical electrons
with internal degree of freedom ( or ``spins'' ) on a finite lattice
$\Lambda =\{x\}$, with sites $x$ in $R$. The internal degrees of freedom
is denoted by $\sigma$, with $\sigma=1, 2, \cdots, \nu$. The
electrons interact with each other
and can also interact with ``classical particles'' (
or f-electrons in the study of rare earth materials, and mixed valence systems
),
as well as with ``impurity spins'' ( e.g. Kondo models ), located
on the vertices of the lattice.

The electrons are described by creation and annihilation
operators $a_{x\sigma}^\dagger, a_{x\sigma}$, which satisfy
the usual anti-commutation relations $\{a_{x\sigma}^\dagger,a_{y\sigma'}\}
=\delta_{xy} \delta_{\sigma\sigma'}, \cdots$;
the classical particles are described by a random variables $\omega_x$,
taking for example the values 1 and 0 to describe the presence or absence
of a particle at the site $x$, and the impurity spin at $x$ is
denoted by $\vec S_x^f$.

The Hamiltonian of physical interest are typically
of the form $H=T+W$, where the kinetic part $T$ is given by
\begin{equation}
T=-\sum_{x,y}\sum_{\sigma,\sigma'} t_{xy}^{\sigma\sigma'} a_{x\sigma}^\dagger
a_{y\sigma'} + \sum_{x,y}\sum_{\sigma,\sigma'}
\tilde t_{xy}^{\sigma\sigma'} a_{x\sigma}^\dagger a_{y\sigma'}
[n_{x,-\sigma} + n_{y,-\sigma'} -\gamma n_{x,-\sigma} n_{y,-\sigma'}],
\end{equation}
where $t_{xy}^{\sigma\sigma}$ is the usual hopping matrix,
$t_{xy}^{\sigma\sigma'}, \sigma\ne \sigma'$, takes into account spin
orbit coupling, $\tilde t_{xy}^{\sigma\sigma'}$, takes into account
bond-charge repulsion (written here for $\nu=2$ with $\sigma=\pm 1$).
The potential part $W$ is given by
\begin{equation}
W=\sum_{x,y}\sum_{\sigma,\sigma'} U_{xy}^{\sigma\sigma'} n_{x\sigma}
n_{y\sigma'} + \sum_{x,y}
J_{x,y} {\vec S_x} \cdot {\vec S_y} + \tilde U \sum_x \omega_x n_x
+ \tilde J \sum_x {\vec S_x^f} \cdot {\vec S_x},
\end{equation}
where $U, J, \tilde U, \tilde J$ are coupling constants
( or interaction potentials ), $n_{x\sigma}=a_{x\sigma}^\dagger a_{x\sigma}$,
and $n_x=\sum_\sigma n_{x\sigma}$.

In this context, the simplest model is the Falicov-Kimball model
which describes spinless electrons ($\nu=1$), hopping
between nearest neighbours and interacting only with
the classical particles,\cite{r:gruber1} i. e.:
\begin{equation}
H_{F.K.}=-t \sum_{<x,y>} a_x^\dagger a_y + \tilde U \sum_x
\omega_x n_x.
\end{equation}
Other typical short range models are\\
$\cdot$ The Hubbard model ($\nu=2, \sigma=\uparrow$ or
$\downarrow$)\cite{r:lieb2},
\begin{equation}
H_H=-\sum_{<x,y>} \sum_\sigma t^{\sigma\sigma} a_{x\sigma}^\dagger
a_{y\sigma} + \tilde U \sum_x n_{x\uparrow} n_{x\downarrow}.
\end{equation}
In the ordinary Hubbard model $t^{\uparrow\uparrow}
=t^{\downarrow\downarrow} =t$. On the other hand,
taking $t^{\uparrow\uparrow} = t$ and $t^{\downarrow\downarrow} =0$,
one recovers the F.K. model with $\sigma=\downarrow$, corresponding
to the classical particles.\\
$\cdot$ The Hubbard model with spin flip for which
$t_{xy}^{\sigma\sigma'}=t$ for nearest neighbours, zero otherwise, and
$\tilde t_{xy}^{\sigma\sigma'}=0$. Introducing the operators\cite{r:holland}
\begin{eqnarray}
A_x^\dagger = {1\over \sqrt 2}
(a_{x\uparrow}^\dagger + a_{x\downarrow}^\dagger)\nonumber\\
B_x^\dagger = {1\over
\sqrt 2} (a_{x\uparrow}^\dagger - a_{x\downarrow}^\dagger),
\end{eqnarray}
one recovers again the F. K. model,
with $B$ describing the classical particles.\\
$\cdot$ The generalized Hubbard model, where
\begin{equation}
H=H_H + V\sum_{<x,y>} n_x n_y + \tilde U \sum_x \omega_x n_x.
\end{equation}
$\cdot$ The Montorsi-Rasetti model, where $\nu=2$ and
$t_{<x,y>}^{\sigma\sigma'} = \tilde t_{<x,y>}^{\sigma\sigma'}=t$
for nearest neighbors.
Introducing the operators $A_x^\dagger$ and $B_x^\dagger$ as above,
one obtain the Hirsch model with $B$ immobile.

One the other hand, several long range models have been
studied, e.g.,\\
$\cdot$ $H=-\sum_{x\ne y} \sum_\sigma t_{xy}
a_{x\sigma}^\dagger a_{y\sigma} + U \sum_x n_x n_x,$
with $t_{xy} = it (-1)^{(x-y)} / d_{xy}$, and
$d_{xy}={L\over \pi} \sin[{\pi(x-y)\over L}]$.\\
$\cdot$ ``Kondo Lattice models", $ H =\sum_{x\ne y}
\sum_\sigma t_{xy} a_{x\sigma}^\dagger a_{y\sigma}
+ J \sum_x {\vec S_x^f} \cdot {\vec S_x}$,
with $t_{xy}=t$.\cite{r:wang5}

In 1969, Lieb and Wu obtained the exact solution
for the 1D Hubbard model using Bethe-ansatz techniques\cite{r:wu}.
Later, motivated by the work of Calogero, Sutherland and
Moser on the integrability of 1D electron in the
continuum with $r^{-2}$ interaction\cite{r:suth72},
Haldane and Shastry
introduced the spin chain with $1/r^2$ interaction\cite{r:haldane,r:shastry}.
In the following,
we shall concentrate on the t-J models with $1/r^2$ interaction and
discuss various properties of these systems.

\section{t-J models}

The t-J models discuss lattice systems of electrons with
hard core, i.e., the wavefunctions for $N_e$ electrons
$\Psi(x_1\sigma_1, \cdots, x_{N_e\sigma_{N_e}}), x_i\in \Lambda,
\sigma_i\in\{1,2,\cdots,\nu\}$ must satisfy the constrain
$\Psi=0$ if $x_i=x_j$ for some pair $(i,j)$.
The Hamiltonian is
\begin{equation}
H_{tJ}=P_G\{ - \sum_{x,y}\sum_{\sigma=1}^\nu t_{xy}
a_{x\sigma}^\dagger a_{y\sigma} +
\sum_{x,y}J_{xy} [ P_{xy} - (1-n_x) (1-n_y)]\} P_G,
\label{eq:tjmodel1}
\end{equation}
where $P_G$ is the Gutzwiller projector onto those states
satisfying the constraint that there is at most one electron
at each site, and $P_{xy}$ is the operator which permutes the spins
of the electrons at sites $x$ and $y$, and which is zero
if the two sites are not both occupied.

Following the usual approach we replace the original
model by a new one with two types
of particles $F$ and $B$ satisfying the constraint that there is exactly
one particle at each site. The $F$ particles are fermions with spins,
described by the operators $f_{x\sigma}^\dagger, f_{x\sigma}$,
$x\in \Lambda, \sigma\in \{1,2,\cdots,\nu\}$, the $B$ particles are spinless
bosons described by $b_x^\dagger,b_x$. The constraint, expressed by
$b_x^\dagger b_x +\sum_\sigma f_{x\sigma}^\dagger f_{x\sigma}=1$,
implies that the wavefunction for $N_e$ fermions and $Q$ bosons
$\phi(x_1\sigma_1,\cdots,x_{N_e}; y_1, \cdots, y_Q)$ is zero, if
$\{x_j\}\bigcup\{y_l\}\ne \Lambda$. In particular, $N_e +Q=N$, the number
of the lattice sites.

The two models are then isomorphic with the obvious mapping
\begin{eqnarray}
&&\Psi (x_1\sigma_1, \cdots, x_{N_e} \sigma_{N_e})
= \phi (x_1\sigma_1, \cdots, x_{N_e}\sigma_{N_e}, y_1, \cdots, y_Q)\nonumber\\
&&a_{x\sigma}^\dagger \rightarrow X_x^{\sigma0}=f_{x\sigma}^\dagger
b_x\nonumber\\
&&a_{x\sigma}\rightarrow X_x^{0\sigma}=b_x^\dagger f_{x\sigma}.
\end{eqnarray}
Furthermore, introducing the operators
\begin{eqnarray}
&&X_x^{\sigma\sigma'}=f_{x\sigma}^\dagger f_{x\sigma'}~~~~~~~~~~~~~~\nonumber\\
&&X_x^{00}=b_x^\dagger b_x,
\end{eqnarray}
the Hamiltonian $H_{tJ}$ is mapped on
\begin{eqnarray}
H&=&P_1\{-\sum_{x,y}\sum_\sigma {1\over 2}
(t_{xy} X_x^{\sigma0} X_y^{0\sigma} -{\bar t_{xy} } X_x^{0\sigma}
X_y^{\sigma0}) +\nonumber\\
&+&\sum_{xy} J_{xy} (\sum_{\sigma,\sigma'}
X_x^{\sigma\sigma'} X_y^{\sigma'\sigma} -X_x^{00} X_y^{00})\} P_1,
\end{eqnarray}
where $P_1$ is the projector onto those states satisfying
those constrains. The t-J model is supersymmetric if
\begin{equation}
t_{xy}={\bar t_{xy}}, ~ J_{xy}={1\over 2} t_{xy},
\end{equation}
and in this case,
\begin{equation}
H=P_1\{-\sum_{x,y} \sum_{a,b} J_{xy} X_x^{ab} X_y^{ba} \theta_b\}P_1,
\label{eq:tjmodel2}
\end{equation}
where $a, b \in \{0,1,\cdots,\nu\}$ and $\theta_0=+1, \theta_\sigma=-1$.
The operators $X_x^{ab}$ generate a superalgebra. This superalgebra
induces superrotations which mix the $F$ and $B$ particles, but leaves
invariant $\sum_a X_x^{aa}=1$. Finally, the supersymmetric t-J model is
invariant under these super-rotations.
The usual $SU(2)$, supersymmetric, short range t-J model on a uniform
lattice, i.e.
\begin{equation}
\nu=2,\cases{t_{xy}=t, J_{xy}=J=t/2, &if $|x-y|=1$, \cr
t_{xy}=J_{xy}=0, &otherwise\cr}
\end{equation}
is integrable, and was solved first by Sutherland in 70's. Following
the works of Haldane and Shastry on the spin chain with $1/r^2$
interaction, Kuramoto and Yokoyama introduced a $SU(2)$,
long range t-J model with\cite{r:kuramoto}
\begin{equation}
\cases{J_{xy}={1\over 2} t_{xy} = 1/d_{xy}^2\cr
d_{xy}={L\over \pi} \sin({\pi (x-y)\over L}).\cr}
\end{equation}
They obtained the ground state wavefunction away from the half-filling.
Later the asymptotic energy spectrum was derived on the assumption
of asymptotic scattering matrix factorization\cite{r:kawakami},
and the system was
identified as a free Luttinger liquid. Generalized Jastrow
wavefunctions for excitations of the system were explicitly
constructed with the representations of down spins and holes by Wang,
Liu and Coleman, and they constructed the full energy spectrum and
thermodynamics of the system in terms of more generalized Jastrow
wavefunctions\cite{r:wang1}, which will be discussed
in the next section.

\section{Supersymmetric t-J model with $1/r^2$ hopping \\and exchange
on the uniform lattice}

With the mapping discussed in Section 3, the eigenvalue equation
$H_{tJ} |\Psi> =E |\Psi>$ for the electron model is mapped onto
the equation $H |\phi> =E |\phi>$ for the F-B model with $H$ given by
Eq.(~\ref{eq:tjmodel2} ) and $|\phi>$ can be written as
\begin{eqnarray}
|\phi> &&=\sum_{\{x\},\{y\}} \sum_{\sigma_1, \cdots,\sigma_{N_e}}
\phi (x_1\sigma_1, \cdots, x_{N_e} \sigma_{N_e}; y_1, \cdots, y_Q)\cdot
\nonumber\\
&&\cdot f_{x_1\sigma_1}^\dagger \cdots f_{x_{N_e}\sigma_{N_e}}^\dagger
b_{y_1}^\dagger \cdots b_{y_Q}^\dagger |0>,
\end{eqnarray}
where $N_e$ is the number of fermions and $Q$ is the number
of bosons on the lattice, $\{x\}\bigcup\{y\}=\Lambda$.
Let us introduce the notation,
\begin{eqnarray}
&&\phi(x_1\sigma_1,\cdots,x_{N_e}\sigma_{N_e}; y_1, \cdots, y_Q)
=\phi(\{q\};\{\sigma\})\nonumber\\
&&(q_1,\cdots,q_N)=(x_1,\cdots,x_{N_e},y_1,\cdots,y_Q), ~~N_e+Q=N.
\end{eqnarray}
The eigenvalue equation for $N_e$ fermions and $Q$ bosons
of the F-B model takes the form
\begin{equation}
-{1\over 2} [\sum_{i\ne j} d^{-2}(q_i-q_j) M_{ij}]\phi(\{q\};\{\sigma\})
=E \phi(\{q\};\{\sigma\}),
\end{equation}
where $M_{ij}$ is the operator which exchange the positions of
particles $i$ and $j$:
\begin{eqnarray}
&&(M_{ij}\phi) (\{q\};\{\sigma\})=\phi(\{q'\};\{\sigma\})\nonumber\\
&&(q_1', \cdots, q_i', \cdots, q_j', \cdots, q_N')=(q_1,\cdots, q_j, \cdots,
q_i, \cdots, q_N).
\end{eqnarray}

At this point, we can use the exchange operator
formalism first introduced by
Polychronakos to exhibit a complete set of constants of motion.
In fact, this method is used by Fowler and Minahan to obtain
the constants of the motion of the spin chain model of Haldane
and Shastry ( to which the t-J model reduces at half-filling ).
Following the idea of Fowler and Minahan\cite{r:fowler},
we define the operators
\begin{equation}
\cases{\pi_j=\pi_j^\dagger = \sum_{k=1(\ne j)}^N
{z_k\over z_j-z_k} M_{jk}, ~~ j = 1,\cdots,N \cr
z_j=\exp(i {2\pi\over N} q_j),\cr}
\end{equation}
and $I_n=\sum_{j=1}^N \pi_j^n, ~~~ n=0,1,2,\cdot, \infty$.
Using the facts that all sites are occupied by exactly
one particle and that the lattice is invariant under translation,
it was shown\cite{r:fowler} that
\begin{eqnarray}
&&[I_n, I_m]=0,\nonumber\\
&&[H, I_n]=0,~~n,m=0,1,2,\cdots.
\end{eqnarray}
Using the properties of the wave function
under permutation it is straightforward to write out all the
constants of the motion
in terms of creation and annihilation operators\cite{r:wang2}.

With the integrability of the model established, the eigenfunctions can be
explicitly constructed. In the $SU(2)$ case, one starts from the fully
polarized up-spin state $|P>$. Let $M$ denote the number of the down
spins and Q the number of the holes, then the wavefunction can be written
as\cite{r:wang1}
\begin{equation}
|\phi>=\sum_{x,y} \phi (x,y) \prod_\alpha S_{x_\alpha}^-
\prod_i h_{y_i}^+|P>,
\end{equation}
where $\phi(x,y)$ is symmetric in $x=(x_1,\cdots,x_M)$, the positions
of the down-spins, and antisymmetric in $y=(y_1,\cdots,y_Q)$, the positions
of the holes; $S_x^-=a_{x\uparrow}^\dagger a_{x\downarrow},
h_y^\dagger=a_{y\uparrow}, S_z=(N-Q)/2-M$. A large class of Jastrow
product eigenstates of uniform motion and spin polarization are
given by\cite{r:wang1}
\begin{eqnarray}
&&\phi(x,y,J_s,J_h) =e^{{2\pi i\over N} (J_s\sum_\alpha x_\alpha
+ J_h\sum_i y_i)}
\phi_0 (x,y)\nonumber\\
&&\phi_0(x,y)=\prod_{\alpha<\beta} d^2(x_\alpha-x_\beta)
\prod_{i<j} d(y_i-y_j) \prod_{\alpha,i} d(x_\alpha-y_i),
\end{eqnarray}
where the quantum numbers $J_s$ and $J_h$ satisfy the following
constrains
$|J_s-N/2|\le N/2-(M-1+Q/2)$,
$|J_h-N/2|\le N/2-(M+Q-1)/2$ and
$|J_h-J_s|\le(M+1)/2$.
Further, the full energy spectrum and thermodynamics can also be
studied by looking at more generalized Jastrow wavefunctions\cite{r:wang1}.

\section{Supersymmetric t-J model with $r^{-2}$ hopping \\and exchange
on a non-uniform lattice}

In this section, we discuss the supersymmetric t-J model
defined on a non-uniform lattice\cite{r:wang3,r:wang4}.
The sites of the lattice
are given by the roots of the Hermit polynomial $H_N(x)$.
It is well known that the Hermit polynomial $H_N(x)$
has $N$ roots, all of which are real and distinct. Therefore
the lattice is well defined. In this case,
$J_{xy}=t_{xy}/2=1/(x-y)^2$. At half-filling, this t-J model
reduces to the spin chain introduced by Polychronakos\cite{r:poly}.
With the super-algebra presentation, and with the permutation
symmetry of the wavefunction, the t-J model Hamiltonian can
take the following form
\begin{equation}
H=-{1\over 2} \sum_{1\le i\ne j\le N} (q_i-q_j)^{-2} M_{ij}.
\end{equation}
One can show that
$[H, I_n]=0, [I_n, I_m]=0$, where
$I_n=\sum_{j=1}^N (a_j^\dagger a_j)^n, a_j^\dagger = i\sum_{k=1}^N
{1\over q_j-q_k} M_{jk} + iq_j, a_j= (a_j^\dagger)^\dagger$,
and $n,m=1,2,\cdots, \infty$. These commutation relations
yield integrability of the system. Having established the
integrability, we have found the ground state of the model
in the subspace of fixed number of holes, and fixed number of
particles of each internal
spin degree $N_\sigma, \sigma=1, 2, \cdots, \nu$\cite{r:wang3}.
\begin{equation}
\phi(x_1\sigma_1, \cdots, x_{N_e}\sigma_{N_e};y_1,\cdots,y_Q)
=\prod_{i<j} (x_i-x_j)^{\delta_{\sigma_i\sigma_j}}
e^{i{\pi\over2} sgn(\sigma_i-\sigma_j)}.
\end{equation}
It has been proved that this wavefunction has eigenenergy
given by\cite{r:wang3}
\begin{equation}
E_0=-{1\over 4} N(N-1) + {1\over 2} \sum_{\sigma}^\nu
\tilde N_\sigma (\tilde N_\sigma-1).
\end{equation}
Furthermore, we have found that the full energy spectrum of the system
in the subspace of fixed number of particles of each internal
spin degree can be written as\cite{r:wang3}
\begin{equation}
E=E_0+ s
\end{equation}
where $s=0,1,2, \cdots, s_{max}$, with an upper bound $s_{max}$
due to the finite size of the Hilbert space.
However, we have been unable to develop a systematic rule to
characterize the degeneracy of the each energy level and to
explain it with the underlying symmetries of the system.

\section{Conclusion and acknowledgement}
In this talk, we have reviewed some recent progresses in
exact solutions of low dimensional electronic systems.
We wish to thank Mo-lin Ge and F. Y. Wu
for the invitation. This work was supported in part by the
Swiss National Science Foundation.



\begin{references}
\bibitem{r:gruber1} C. Gruber and N. Macris,
{\it The Falicov-Kimball: review of rigorous
results}, (1995). References therein.
\bibitem{r:wu} E. H. Lieb and F. Y. Wu,
Phys. Rev. Lett. {\bf 20}, 1445 (1968).
C. N. Yang, Phys. Rev. Lett.
{\bf 19}, 1312 (1967).
\bibitem{r:holland} K. Michielson, Int. J. Mod. Phys. {\bf B 7},
2571 (1993).
\bibitem{r:lieb2} E. H. Lieb, {\it The Hubbard model: some rigorous results
and open problems}, in {\it Proceedings of the Conference ``
Advances in Dynamical Systems and Quantum Physics''}
(World Scientific, Singapore, 1995).
\bibitem{r:haldane}
F. D. M. Haldane, Phys. Rev. Lett. {\bf 60}, 635 (1988).
\bibitem{r:shastry} S. Shastry, Phys. Rev. Lett. {\bf 60}, 639 (1988).
\bibitem{r:kuramoto} Y. Kuramoto and H. Yokoyama, Phys. Rev. Lett. {\bf
67}, 1338 (1991).
\bibitem{r:kawakami} N. Kawakami, Phys. Rev. B {\bf 45}, 7525 (1992).
\bibitem{r:wang1} D. F. Wang, James T. Liu and P. Coleman,
Phys. Rev. {\bf B 46}, 6639 (1992).
\bibitem{r:poly} A. P. Polychronakos,
Phys. Rev. Lett. {\bf 70}, 2329 (1993).
\bibitem{r:fowler} M. Fowler and J. Minahan, Phys. Rev. Lett.
{\bf 70}, 2325 (1993).
\bibitem{r:wang2} D. F. Wang and C. Gruber, Phys.
Rev. {\bf B 49}, 15712 (1994).
\bibitem{r:wang3}
C. Gruber and D. F. Wang, Phys. Rev. {\bf B 50}, 3103 (1994).
\bibitem{r:wang4}
C. Gruber and D. F. Wang, invited talk at King's College,
May 1994.
\bibitem{r:suth72} B. Sutherland, Phys. Rev. A {\bf 5}, 1372 (1972),
Phys. Rev. A {\bf 4}, 2019 (1971).
\bibitem{r:wang5} D. F. Wang and C. Gruber, Phys. Rev. {\bf B 51}, 7476 (1995).
\end{references}
\end{document}